\documentstyle[12pt]{article}
\def\be{\begin{equation}}
\def\ee{\end{equation}}
\def\bea{\begin{eqnarray}}
\def\eea{\end{eqnarray}}

\begin{document}
\begin{flushright}
hep-th/9909167 \\
IPM/P-99/015
\end{flushright}
\begin{center}
{\Large{\bf Branes with Back-ground Fields in Boundary State Formalism }}                  
										 
\vskip .5cm   
{\large H. Arfaei and D. Kamani}
\vskip .1cm
 {\it Institute for Studies in Theoretical Physics and 
Mathematics
\\ Tehran P.O.Box: 19395-5531, Iran}\\
{\it and}\\
{\it  Department of Physics, Sharif University of Technology
P.O.Box 11365-9161}\\
{\sl e-mail: arfaei@theory.ipm.ac.ir\\ 
e-mail: kamani@netware2.ipm.ac.ir}
\\
\end{center}

\begin{abstract} 
Interaction of branes in presence of internal gauge fields  
is considered by using the boundary state formalism . 
This approach enables us to consider the problems that are not easily 
accessible to the canonical approach via open strings . 
The effects of compactification of some of the 
dimensions on tori are also discussed .
 Also we study the massless state contribution on this interaction .
\end{abstract} 
\vskip .5cm

\newpage
\section{Introduction}
 
$D_{p}$-branes with non zero back ground internal gauge fields,
 or equivalently D-branes in $B_{\mu\nu}$ fields and $U(1)$
gauge fields $A_{\alpha}$ have shown several interesting 
properties \cite{1,2,
3,4,5,6,7,8}. On the one hand they describe bound states of 
$D_{p}$ and $D_{p-2}$-branes \cite{4} or $D_{p}$-branes and fundamental strings  
\cite{3} and on the other hand they give rise to SYM theories on a non 
 commutative space \cite{9} . Also the study of the physics of (n,1) 
bound states \cite{5,6}, creation of fundamental string between   
$D_4$-branes \cite{7}, loop corrections to superstring equations   
of motion \cite{8}, scattering of closed strings from D-branes 
\cite{1} and dynamics and interactions of D-brane solitons \cite{2}  
are another applications of $D_p$-branes in non zero back ground  
gauge fields . A useful tool for describing  $D_p$-branes and 
their interactions (specially  
in non zero back ground gauge fields) is the boundary state formalism  
 \cite{1,10,11,12,13,14,15} . 
  Simply the overlap of boundary states through the closed string propagator 
 give us
 the amplitude of interaction of branes . By introducing back ground
 fields $B_{\mu\nu}$ and a $U(1)$ gauge field 
 $A_{\alpha}$ (which lives in D-brane) to the 
 string $\sigma$-model action one obtains mixed boundary 
 conditions for the strings emitted by the branes
 . The method of boundary states enables us to study the  
 interaction of branes with different dimensions and different 
 internal fields some of which are not easily accessible by 
 canonical formalism \cite{4}.
   
   We shall also consider compactification of certain dimensions on tori .
   The states emitted from the branes which are wrapped around compact 
  directions with internal back ground fields 
  turn out to be dominantly along a certain direction 
  not perpendicular to the brane which specified by its windings . 
  These windings,
  around the compact directions are correlated 
  with their momenta 
  along the brane . This is in contrast to the case of pure D-brane
  where the closed strings only have momentum perpendicular to the 
  brane .
  
  In section 2 we write and solve the equations of boundary states 
  . In section 3 we use the result of section 2
  to calculate the interaction of two branes of arbitrary dimensions 
  ${p_1}$ and ${p_2}$ with different internal ${\cal{F}}$ fields. 
  We also find the result when part of the space is compactified 
  on a torus. We shall also show that our result reduces to 
  that of the known cases such as 
  zero ${\cal{F}}$ fields, and non-compact spacetime .
  Finally the contribution of the massless states on the amplitude 
  is extracted and thus interaction strength is be obtained . 
  
  Since compactification effects on the interaction of the  
  branes do not depend on the fermions,  
   we will consider only the bosonic string .
  In this article we denote a brane in the back-ground of internal fields 
  by ``$m_p$ -brane'', $i.e.$ a ``mixed brane'' with dimension 
  ``$p$''.
\section{Boundary state} 

  We begin with a $\sigma$-model action containing $B_{\mu\nu}$ field and two  
boundary terms \cite{16} corresponding to the two  $m_{p_{1}}$ and 
$m_{p_{2}}$-branes gauge fields .

\bea
S=-\frac{1}{4\pi\alpha'}\int_{\Sigma}d^{2}\sigma
\bigg{(} \sqrt{-g}g^{ab}G_{\mu\nu}
\partial_{a}X^{\mu}\partial_{b}X^{\nu}+\epsilon^{ab}B_{\mu\nu}\partial
_{a}X^{\mu}\partial_{b}X^{\nu}\bigg{)}
\nonumber\\
-\frac{1}{2\pi\alpha'}\int_{({\partial\Sigma})_{1}}
d\sigma A^{(1)}_{\alpha_1}\partial_{\sigma}X^{\alpha_{1}}
+\frac{1}{2\pi\alpha'}\int_{({\partial\Sigma})_{2}}d\sigma A^{(2)}_{\alpha_2}
\partial_{\sigma}X^{\alpha_{2}}                                    
\eea

where $\Sigma$ is the world sheet of the closed string exchanged
between the branes . $(\partial\Sigma)
_{1}$ and $(\partial\Sigma)_{2}$ are two boundaries of this world sheet, 
which are at $\tau=0$ and $\tau=\tau_{0}$ respectively .
$A^{(1)}_{\alpha_1}$ and $A^{(2)}_{\alpha_{2}}$ are $U(1)$ gauge 
fields that live in $m_{p_1}$ and $m_{p_2}$-branes . The sets 
$\{{\alpha_1}\}$ and $\{{\alpha_2}\}$ 
 specify the directions on the $m_{p_1}$ and $m_{p_2}$-world branes .
 $G_{\mu\nu}$ and $B_{\mu\nu}$ are usual back-ground fields .
 Vanishing the variation of this action with respect 
 to $X^{\mu}(\sigma,\tau)$ gives
the equation of motion of $X^{\mu}(\sigma,\tau)$ and boundary state equations.

Taking the $B_{\mu\nu}(X)$ and $G_{\mu\nu}(X)$ to be constant
fields, thus the boundary states equations become 
\bea
&~&\bigg{(} \partial_{\tau}X^{\alpha_1}+{\cal{F}}^
{\alpha_1}_{(1)\;\beta_1}\partial_
{\sigma}X^{\beta_1} +B^{\alpha_1}\;_{i_1} \partial_{\sigma} X^{i_1}
\bigg{)}_{\tau=0}{\mid}B^1_x \rangle = 0       
\nonumber\\
&~&{({\delta}X^{i_ {1}})}_{\tau=0}{\mid}B^1_x \rangle = 0
\nonumber\\
&~&\bigg{(}\partial_{\tau}X^{\alpha_2}+
{\cal{F}}^{\alpha_2}_{(2)\;\beta_2}\partial_{\sigma} 
X^{\beta_2} +B^{\alpha_2}\;_{i_2}\partial_{\sigma}X^{i_2}
\bigg{)}_{\tau=\tau_0}{\mid}B^2_x \rangle = 0
\nonumber\\
&~&{({\delta}X^{i_{2}})}_{\tau=\tau_0}{\mid}B^2_x \rangle = 0
\eea
where $\{i_1\}$ and $\{i_2\}$ show the directions perpendicular 
to $m_{p_1}$ and $m_{p_2}$-world branes and total ``field strengths'' are  
\bea
{\cal{F}}_{(1)\alpha_1\beta_1}=B_{\alpha_1\beta_1}-A^{(1)}_{[
\alpha_1,\beta_1]}
\nonumber\\
{\cal{F}}_{(2)\alpha_2\beta_2}=B_{\alpha_2\beta_2}-A^{(2)}_{[
\alpha_2,\beta_2]}.
\eea
Furthermore if we take $m_{p_1}$ and $m_{p_2}$-branes to be at positions 
$\{y^{i_{1}}_1\}$ and  $\{y^{i_{2}}_2\}$  respectively, we have to impose  
\bea 
&~&[X^{i_1}(\sigma,\tau)-y^{i_1}_{1}]_{\tau=0}{\mid}B^1_x \rangle = 0
\nonumber\\
&~&[X^{i_2}(\sigma,\tau)-y^{i_2}_{2}]_{\tau_0}{\mid}B^2_x \rangle = 0
\eea   
The solution to the equation of motion is
\bea
X^{\mu}(\sigma,\tau)= x^{\mu}+2{\alpha'}p^{\mu}\tau+2L^{\mu}\sigma+
 \frac{i}{2}\sqrt{2\alpha'}\sum_{m\neq 0}\frac{1}{m}(\alpha^{\mu}_{m}
 e^{-2im(\tau-
 \sigma)}+\tilde\alpha^{\mu}_{m}e^{-2im(\tau+\sigma)})
\eea
where $L^{\mu}$ is zero for non-compact directions but can be non-zero
for the components along the compact directions, in which 
\bea
&~&L^{\mu}=N^{\mu}R^{\mu} \;\;\;\;,\;\;\;\; (N^{\mu} \in Z)\;\;\;\;
( no\; sum\; on\; \mu )
\eea
where $R^{\mu}$ is the radius of compactification in the compact direction 
$X^{\mu}$ . Also for the momenta in such directions we have 
\bea
&~&p^{\mu}=\frac{M^{\mu}}{R^{\mu}} \;\;\;\;,\;\;\;\; (M^{\mu} \in Z)
\eea
 $N^{\mu}$ is the winding number and $M^{\mu}$ is the momentum number of the
 closed string state . 
 Imposing the boundary conditions on the solution (5) as operator equation, 
 we obtain  the boundary 
 state equations that, for the second brane take the form
\bea
\bigg{(}p^{\alpha_2}+\frac{1}{\alpha'}{\cal{F}}^{\alpha_2}
_{(2)\;\beta_2} L^{\beta_{2}}
\bigg{)}\mid B^2_x,\tau_{0} \rangle = 0
\eea                                           
\bea
\bigg{(}(1-{\cal{F}}_{2})^{\alpha_{2}}_{\;\;\;\beta_{2}}
\alpha^{\beta_2}_{n}e^{-2in\tau_{0}}+
(1+{\cal{F}}_{2})^{\alpha_2}_{\;\;\;\beta_2}
\tilde\alpha^{\beta_2}_{-n}e^{2in\tau_0}
\bigg{)}\mid B^2_x,\tau_0 \rangle = 0
\eea
\bea
\bigg{(}\alpha^{i_2}_{n}e^{-2in\tau_{0}}-\tilde\alpha^{i_2}
_{-n}e^{2in{\tau_0}}
\bigg{)}\mid B^2_x,\tau_{0} \rangle = 0
\eea
\bea
(x^{i_2}+2{\alpha'}p^{i_2}\tau_{0}-y^{i_2}_{2})
{\mid}B^2_x,\tau_{0} \rangle = 0
\eea
\bea                
L^{i_2}\mid B^2_x,\tau_{0} \rangle=0
\eea
where $\tau_0$ is the $\tau$ variable on the boundary of the closed
string world sheet. 
 If the direction $X^{i_2}$ is non-compact we have $L^{i_2}=0$, and if this 
direction is compact then equation (12) implies that $L^{i_2}=0$
, $i.e.$ the closed string cannot 
wind around directions perpendicular to the brane .
Equation (8) also implies that the closed string 
 can have a momentum in the world brane directions 
 . This is in contrast to the non compact 
 case where the momentum components along the brane directions 
 are zero . As equation (8) shows the momentum and the windings of   
 the emitted string are proportional . Physically it means that when
 the back ground ${\cal{F}}$ is turned on and brane is wrapped 
 around the compact directions, the closed string that is emitted 
 (absorbed) from (by) the   
 brane must have its momentum in one direction of brane proportional 
 to its winding numbers in the other compact directions of brane .
 According to this equation, 
 closed string momentum components along both compact and 
 non compact directions of the brane is 
 quantized. It is rather surprising that the internal momentum even in  
 non compact directions of the brane is non zero and quantized . It is worth
 noting that although this momentum is non zero, it is not an independent 
 quantum number. This is the effect of internal 
 gauge fields and compactification.
 For the compact directions of brane also we have quantized momenta 
 $p^{{\alpha}_c} = \frac{M^{\alpha_c}}{R^{\alpha_c}}$ ,  
 therefore
 \bea
 M^{\alpha_c} = -\frac{1}{\alpha'}\sum_{\beta_c}{\cal{F}}^
 {\alpha_c}\;_{\beta_c}
 N^{\beta_c}R^{\beta_c}R^{\alpha_c}
 \eea
 In terms of $\phi^{\alpha_c}\;_{\beta_c} = 4\pi^2{\cal{F}}^{\alpha_c}
 \;_{\beta_c}R^{\alpha_c}R^{\beta_c}$, the flux through the 2-cycle
 $\{\alpha_c\beta_c\}$ we have,
 \bea
 M^{\alpha_c} = -\frac{1}{4\pi^2\alpha'}\sum_{\beta_c}\;\phi^{\alpha_c}\;
 _{\beta_c}N^{\beta_c}
 \eea
 As $M^{\alpha_c}$ and $N^{\beta_c}$ are integers we can look at this 
 relation as a constraint on the flux allowing a closed string with 
 momentum numbers $\{M^{\alpha_c}\}$ and winding numbers 
 $\{N^{\alpha_c}\}$ . 
 $M^{\alpha_c}$ and $N^{\beta_c}$
 being integers restrict $-\frac{1}{4\pi^2\alpha'}\phi^
 {\alpha_c}\;_{\beta_c}$ 
 to be rational . On the other hand for a given flux the 
 relation between momentum numbers 
 and winding numbers of closed strings specifies which string lives in 
 $\mid B \rangle$ and can be emitted. Similar effect for the open 
 strings ending on the branes is observed \cite{9} .

Now we solve the boundary state equations . Equation (11) 
and (12) have the solution
\bea
 e^{i\alpha'\tau_{0}\sum_{i_2}(p^{i_2}_{op})^{2}}
 {\delta^{(d-p_{2}-1)}}(x^{i_2}-
 y^{i_2}_{2}){\prod_{i_2}}\mid p^{i_2}_{L}=p^{i_2}_{R}=0 \rangle 
\eea
 solution to the equation (8) is, 
\bea
 \sum_{\{p^{\alpha_2}\}}\prod_{\alpha_2}\mid p^{\alpha_2} \rangle
 \eea
 where 
\bea
 p^{\alpha_2}=-\frac{1}{\alpha'}\;{\cal{F}}^{\alpha_2}_
 {(2)\;\;\beta_2c}{\ell}^{\beta_{2c}}
\eea
 $\beta_{2c}$ is index for compact directions of the set 
 $\{X^{\alpha_2}\}$, 
and ${\ell}^{\beta_{2c}}$ is $N^{\beta_{2c}}R^{\beta_{2c}}$ .

The solution of the oscillator parts of boundary state equations is 
\bea
 e^{-\sum_{m=1}^{\infty}\frac{1}{m}e^{4im\tau_0}\alpha^{\mu}_{-m}S^{(2)}_
 {\mu\nu}{\tilde\alpha}^{\nu}_{-m}}\mid 0 \rangle
\eea
  where the matrix $S^{\mu}_{(2)\;\;\nu}$ is 
\bea
S^{\mu}_{(2)\;\;\nu}=(Q^{\alpha_2}_{(2)\;\;\beta_2}\;,\;-\delta^{i_2}
\;_{j_2})
\eea
\bea
Q_2 \equiv {(1-{\cal{F}}_{2})^{-1}}(1+{\cal{F}}_{2})
\eea
 Note that since ${\cal{F}}_{2}$ is
 antisymmetric $Q_2$ is an orthogonal matrix .
 Putting all these together, we obtain the boundary state  
\bea
 \mid B^2_x , \tau_{0} \rangle = \sum_{\{{p^{\alpha_2}}\}}
 \mid B^2_x , \tau_0\;,\;p^{\alpha_2} \rangle
\eea
 The summation is over all possible momenta, which can equivalently
 be written as sum over winding 
 numbers $\{N^{\alpha_{2c}}\}$ using the relation (17) and
\bea  
\mid B^2_x , \tau_0 , p^{\alpha_2} \rangle 
= \frac{T_{p_2}}{2}
\sqrt{det(1-{\cal{F}}_2)} \;e^{i\alpha'\tau_0\sum_{i_2}(p^{i_2}_{op})^2
}\delta^{(d-p_2-1)}(x^{i_2}-y^{i_2}_2) 
\nonumber\\
\times e^{
-\sum_{m=1}^\infty\frac{1}{m}e^{4im\tau_0}
\alpha^\mu_{-m}S^{(2)}_{\mu\nu}\tilde{\alpha}^{\nu}_{-m}
}
\mid 0 \rangle
\prod_{i_2}\mid p^{i_2}_L = p^{i_2}_R = 0 \rangle  
\prod_{\alpha_2}\mid p^{\alpha_2} \rangle
\eea  
The constant $T_{p_2}$ is the tension of $D_{p_2}$-brane
 and is derived in \cite{1} and \cite{14}. 
 The origin of the factor $\sqrt{det(1-{\cal{F}}_{2})}$ is in 
 the path integral with boundary action \cite{8,17,18} .
 The ghost part of boundary state is independent of ${\cal{F}}_2$ and 
 is given by the same expression as for ${\cal{F}}_2=0$
\bea
  \mid B_{gh}, \tau_{0} \rangle = e^{\sum_{m=1}^{\infty}{e^{4im\tau_0}}
  (b_{-m}{\tilde{c}}_{-m}+c_{-m}{\tilde{b}}_{-m})}\mid Z \rangle
\eea
  where $\mid Z \rangle$ is the appropriate 
  ghost vacuum \cite{19} which can be 
  written as 
\bea
  \mid Z \rangle  = (c_{0}-{\tilde{c}}_{0})
  \mid \downarrow\downarrow \rangle
\eea
\section{Interaction between two mixed branes}
 Now we can calculate the overlap of the two boundary states to obtain
 the interaction amplitude of branes.
As compactification effects do not depend on fermions and superghosts
therefore we discuss only on the bosonic case . Complete boundary state is
\bea
\mid B \rangle = \mid B_{x} \rangle \mid B_{gh} \rangle
\eea
 These two mixed branes simply interact via exchange of closed strings, so
 that the amplitude is given by
 \bea
 {\cal{A}} = \langle B_1\mid D \mid B_2,\tau_0 =0 \rangle
 \eea
 where ``$D$'' is the closed string propagator \cite{11}. The calculation 
 is straight forward but tedious.

Here we only give the final result;
\bea
{\cal{A}} &=& \frac{T_{p_1}T_{p_2}}{4(2\pi)^{d_i}}\alpha'
\sqrt{det(1-{\cal{F}}_1)det(1-{\cal{F}}_2)}
\int_0^{\infty} dt \bigg{\{}  
\nonumber\\
&~&\times e^{4at} \prod_{n=1}^\infty \bigg{(}[ det(1-S_1S_2^T 
e^{-4nt})]^{-1}(1-e^{-4nt})^2 \bigg{)}
\nonumber\\
&~&\times \bigg( \sqrt{\frac{\pi}{\alpha't}} \; \bigg)^{d_{i_n}}
e^{ -\frac{1}{4\alpha't}\sum_{i_n}(y^{i_n}_1-y^{i_n}_2)^2 }
\prod_{i_c}\Theta_3 \bigg( \frac{y^{i_c}_1 - y^{i_c}_2 }{2\pi R_{i_c}} \mid 
\frac{i\alpha't}{\pi (R_{i_c})^2}\bigg)
\nonumber\\
&~& \times \sum_{\{N^{u_c}\}} \bigg{[} (2\pi)^{d_u}
[\prod_{u}\delta(p^u_1 , p^u_2)]
\; exp [\frac{i}{\alpha'}{\ell}^{u_c}
({\cal{F}}^{\alpha'_1}_{(1)\;\;u_c} y^{\alpha'_1}_{2}-
{\cal{F}}^{\alpha'_2}_{(2)\;\;u_c}y^{\alpha'_2}_{1})]  
\nonumber\\
&~&\times exp [-\frac{t}{\alpha'}{\ell}^{u_c}{\ell}^{v_c}
(\eta_{u_cv_c}+{\cal{F}}^{u}_{(1)\;\;u_c}{\cal{F}}_{(2)\;\;uv_c}+
{\cal{F}}^{\alpha'_1}_{(1)\;\;u_c}{\cal{F}}^{\alpha'_1}_{(1)\;\;v_c}+
{\cal{F}}^{\alpha'_2}_{(2)\;\;u_c}{\cal{F}}^{\alpha'_2}_{(2)\;\;v_c})] 
\bigg{]} \;\bigg{\}}
\eea
where ``$a$'' is a constant, depends on spacetime dimension $a=(d-2)/24$
. The notation used in the above calculation is given below

\noindent $\{i\} \equiv$ Indices for directions 
perpendicular to the both mixed branes .\\
$\{u\} \equiv$ Indices for directions along on the both mixed branes .        \\
$\{\alpha'_{1}\} \equiv$ Indices for directions along on the $m_{p_1}$ and 
	     perpendicular to the $m_{p_2}$-branes .\\
$\{\alpha'_{2}\} \equiv$ Indices for directions along on the $m_{p_2}$ and
perpendicular to the $m_{p_1}$-branes . \\
These sets with $\{i_1\}$ and $\{i_2\}$ have set notations as 
\bea
\{i_{1}\} = \{i\} \bigcup \{\alpha'_{2}\} \nonumber\\
\{i_{2}\} = \{i\} \bigcup \{\alpha'_{1}\} \nonumber\\
\{\alpha_{1}\} = \{u\} \bigcup \{\alpha'_{1}\} \nonumber\\
\{\alpha_{2}\} = \{u\} \bigcup \{\alpha'_{2}\} \nonumber\\
\{\mu\} = \{i_{1}\} \bigcup \{\alpha_{1}\} = \{i_{2}\} \bigcup 
 \{\alpha_{2}\} .
\eea

 In this amplitude $p^{u}_{1} = -\frac{1}{\alpha'}
  {\cal{F}}^{u}_{(1)\;\;v_c}
 N^{v_c}R^{v_c}$ and $p^{u}_{2} = -\frac{1}{\alpha'}
 {\cal{F}}^{u}_{(2)\;\;v_c}N^{v_c}
 R^{v_c}$ . 
Indices $\{u_{c} , v_{c} ,... \}$ show those directions of $\{X^{u}\}$ which
are compact . $d_{u}$ is dimension of $\{X^{u}\}$ and $d_i$ is dimension 
of $\{X^i\}$. Also $\{i_{n}\}$
is non-compact part of $\{i\}$, and $\{i_{c}\}$ is compact part of 
$\{i\}$ region . ${\ell}^{u_{c}}$ as previous is $N^{u_{c}}R^{u_c}$.  
This amplitude as expected is symmetric under the exchange of 
indices ``1'' and ``2'' (see (26)). Although in this article we are        
dealing only with bosonic string, it is wroth noting that similar 
consideration concerning the effects of compactification works for the 
superstring case since these effects are independent of the fermions .
 
 We can write this amplitude in another form in which common world volume
 of the world branes explicitly appears . 
 In the sum over $\{N^{u_c}\}$ one term is obtained 
 for $N^{u_c} = 0 $ for all $u_c$, therefore
 $p^{u}_{1} = p^{u}_{2} = 0 $ . Furthermore for given fields 
 and radii of compactification  there may be other sets $\{N^{u_c}_{s}\}$ in  
  which equality ${\cal{F}}^u_{(1) v_c}N^{v_c}_s R^{v_c} = 
  {\cal{F}}^u_{(2)v_c} N^{v_c}_s R^{v_c} $ holds
  (one possibility is that $({\cal{F}}_1 -{\cal{F}}_2)
  ^u\;_{v_c}R_{v_c}$ be rational ), then amplitude takes 
  the form  
\bea 
{\cal{A}} &=&  
\frac{T_{p_1}T_{p_2}}{4(2\pi)^{d_i}}\alpha'
V_u\sqrt{det(1-{\cal{F}}_{1})det(1-{\cal{F}}_{2})}
\int_0^\infty dt {\bigg \{} {\bigg (}\sqrt{
\frac{\pi}{\alpha't} }\;{\bigg)}^{d_{i_n}} 
e^{-\frac{1}{4\alpha't}\sum_{i_n}(y^{i_n}_1-y^{i_n}_2)^2} 
\nonumber\\
&~&\times 
\; e^{4at}\prod_{n=1}^\infty \bigg{(}[det(1-S_{1}S_{2}^T e^{-4nt})]^{-1}
(1-e^{-4nt})^2 \bigg{)} 
\prod_{i_c}\Theta_3 {\bigg(}   
\frac{y^{i_c}_1-y^{i_c}_2}{2\pi R_{i_c}} \mid
\frac{i\alpha't}{\pi (R_{i_c})^2}{\bigg )}
\nonumber\\
&~& \times  {\bigg [} 1+ \sum_s \bigg{(}
exp[-\frac{t}{\alpha'}{\ell}^{u_c}_s\ell^{v_c}_s(\eta_{u_c v_c}
+{\cal{F}}^u_{(1) u_c}{\cal{F}}_{(2) uv_c}
+{\cal{F}}^{\alpha'_1}_{(1) u_c}{\cal{F}}^{\alpha'_1}_{(1) v_c}+
{\cal{F}}^{\alpha'_2}
_{(2) u_c}{\cal{F}}^{\alpha'_2}_{(2) v_c})]  
\nonumber\\
&~&  \times exp[\frac{i}{\alpha'}{\ell}^{u_c}_s({\cal{F}}^
{\alpha'_1}_{(1) u_c}
y^{\alpha'_1}_{2}-{\cal{F}}^{\alpha'_2}_{(2) u_c}y^{\alpha'_2}_1) 
] \bigg{)} {\bigg ]}\; {\bigg \}}
\eea
where $V_{u}$ is the common world volume of the two mixed branes  
and ${\ell}^{u_{c}}_{s} = N^{u_{c}}_{s}R^{u_{c}} $.
If there are no sets ${\{N^{u_{c}}_{s}\}}$ then the expression in the 
bracket is 1.
Note that for parallel mixed branes with the same dimension those terms 
which contain $\alpha'_1$ and $\alpha'_2$ disappear .

The effects of compactification are in the factors $\Theta_3$ and 
the last bracket which reduce to 1 for the non compact case . 
Therefore the amplitude for non compact spacetime is the remainder 
part with the change $i_n \rightarrow i$ .

Specially consider two parallel mixed branes  with the same dimension   
$p$ in non compact spacetime with dimension $d$ . In this case $V_u$ is 
$V_{p+1}$ . Furthermore consider ${\cal{F}}_1 = 
{\cal{F}}_2 \equiv {\cal{F}}$, so we have ${(S_{1}S_{2}^T)}^
{\mu}\;_{\nu} 
= {\delta}^{\mu}\;_{\nu}$ 
therefore 
\bea
{\cal{A}} &=& \frac{T^2_p}{4(2\pi)^{d-p-1}}\alpha'
V_{p+1}det(1-{\cal{F}}) \int_0^\infty dt \bigg\{
\bigg(\frac{\pi}{\alpha't}\bigg)^{(d-p-1)/2}
\nonumber\\
&~&\times e^{-\frac{1}{4\alpha't}\sum_{i}(y^i_1-y^i_2)^2}
e^{(d-2)t/6}\prod_{n=1}^\infty (1-e^{-4nt})^{-d+2} \bigg\} 
\eea
Then for $T_p = \frac{\sqrt{\pi}}{2^{(d-10)/4}}(4{\pi}^{2}\alpha')^{  
(d-2p-4)/4}$ after a 
transformation $t\rightarrow {\pi}/{2t}$, equation (30) can
be interpreted as the one-loop free energy of an open string whose ends
are fixed on these parallel branes \cite{20} .

The formula (27) reduces to the previous known results on the   
interaction of mixed branes, for example those considered in \cite{3,4}
. In the previous methods each particular relative configuration of
two branes needs to be treated individually, which may or may not yield to 
the canonical quantization. In contrast,
the method of boundary state can be used for any 
two arbitrary branes. There is also another advantage in the boundary 
state method . In the canonical method when the magnetic part of 
${\cal{F}}$ is different from zero, the space becomes non commutative
. In the present method we can go around this problem and all cases
can be handled in the same way . In the following we consider an 
example using the general formula .

{\bf Example:}

Consider two parallel $m_2$-branes along ($X^1 , X^2$), with non zero 
components of fields 
${\cal{F}}_{(1)\;01} = E$ and ${\cal{F}}_{(2)\;01} = E'$ . 
In this case non zero components of momenta along the world volume are
$p^0_1 = \frac{1}{\alpha'} nER_1 $ and $p^0_2 = \frac{1}{\alpha'}
nE'R_1$, where $n$ is winding number of a given closed string states
around the $X^1$-direction .
Therefore the interaction amplitude becomes 
\bea
{\cal{A}} &=& \frac{T^2_2}{4(2\pi)^{d-3}}\alpha'
V_3 \sqrt{(1-E^2)(1-E'^2)}\;\int_{0}^{\infty}
dt{\bigg\{}e^{4at}
\prod_{n=1}^{\infty}{\bigg[}(1-e^{-4nt})^{4-d}
\nonumber\\
&~&\times{\bigg(}1 - \frac{(1+E)
(1-E')}{(1-E)(1+E')}e^{-4nt}{\bigg)}^{-1}
{\bigg(}1 - \frac{(1-E)(1+E')}{(1+E)(1-E')}e^{-4nt}{\bigg)}^{-1}
{\bigg]}
\nonumber\\
&~&\times{\bigg(}\sqrt{\frac{\pi}{\alpha't}}\;{\bigg)}^{d_{i_n}}
e^{-\frac{1}{4\alpha't}\sum_{i_n}(y^{i_n}_1-y^{i_n}_2)^2}
\prod_{i_c}\Theta_3{\bigg(}\frac{y^{i_c}_1-y^{i_c}_2}
{2{\pi}R_{i_c}} \mid \frac{i\alpha't}{{\pi}(R_{i_c})^2}{\bigg)}
\nonumber\\
&~&\times \theta (E, E', t, R_1) 
\Theta_3(0 \mid itR^2_2/\pi\alpha'){\bigg\}}       
\eea
where $V_3$ is the world volume which can be written as 
$(2{\pi}R_1)(2{\pi}R_2)L$ in which $R_1$ and $R_2$ are radii of 
compactification of $X^1$ and $X^2$ directions respectively and
$L$ is the infinite time length . 
The function $\theta$ is defined by infinite series 
\bea
\theta(E, E',t,R_1) = \frac{2\pi}{L} \sum_{n=-\infty} ^{\infty}
\delta [nR_1(E-E')/\alpha']
\; e^{-t(1-E^2)R^{2}_1n^2/\alpha'}
\eea
For the case in which $E = E'$, this becomes
$\theta(E, E, t, R_1) = \Theta_3 \bigg{(} 0 \mid 
\frac{it(1-E^2)R^2_1}{\pi \alpha'}
\bigg{)} $, and for $E \neq E'$ it is equal to 1 . 

For the non compact case the effect of non zero fields $E=E'$ appear only
in the modification of tension of each brane 
by a factor $\sqrt{1-E^2}$ . On the other hand when $X^1$-direction 
is compact and $E=E'$, the back-ground fields appear not only in the
tensions but also in an extra factor $\Theta_3 \bigg{(} 0 \mid
it(1-E^2)R^2_1/(\pi \alpha') \bigg{)}$ in the above expression .
\section{Contribution of massless states ($B_{\mu\nu}$, 
$G_{\mu\nu}$, $\phi$) on
amplitude}

  For distant branes only massless states 
  have a considerable contribution
on the amplitude. Therefore we will restrict our 
attention to the long-range   
force. We know that the metric, antisymmetric tensor and 
dilaton states have zero winding numbers and zero momentum numbers,
therefore from the general expression of the interaction amplitude (27),
only the term with $N^{u_c}=0$ ( for all $u_c$ ) corresponds to this three
massless states . We also must calculate the following limit
\bea
\Omega \equiv \lim _{q \rightarrow 0}\; \frac{1}{q}\; \prod_{n=1}
^{\infty} \bigg{[} (1-q^n)^2 [det(1-Sq^n)]^{-1} \bigg{]}
\eea
where $q=e^{-4t}$ and $S=S_1S^T_2$, ( note that we imposed $d=26$ ). 
For a matrix $A$ we have $det A = e^{Tr[ln A]}$  therefore
\bea
\Omega = \lim_{q \rightarrow 0}\; \frac{1}{q}\; \prod_{n=1} ^{\infty}
\bigg{[} exp \bigg{(}\frac{q^n}{n(1-q^n)}[Tr(S^n)-2]\bigg{)} \bigg{]}
= \lim_{q \rightarrow 0}\; \frac{1}{q}\; +\; ( TrS\; - \; 2 )
\eea
The leading divergence is from the tachyon and we put away it, therefore
we find
\bea
{\cal{A}}_0 &=& \frac{T_{p_1}T_{p_2}}{4(2\pi)^{d_i}}\alpha'
V_u\sqrt{det(1-{\cal{F}}_1)
det(1-{\cal{F}}_2)}\;\;[Tr(S_1S_2^T)-2]
\nonumber\\
&~&\times\int_{0}^{\infty}dt{\bigg[}{\bigg(}\sqrt{\frac{\pi}{\alpha't}}
\;{\bigg)}^{d_{i_n}}
e^{-\frac{1}{4\alpha't}\sum_{i_n}(y^{i_n}_1-y^{i_n}_2)^2}\prod_{i_c}
{\Theta_3}{\bigg(}\frac{y^{i_c}_1-y^{i_c}_2}{2{\pi}R_{i_c}} \mid \frac
{i\alpha't}{{\pi}(R_{i_c})^2}{\bigg)}{\bigg]}
\eea
Note that the factor 2 in $[Tr(S_1S^{T}_2)-2]$ 
is purely due to the ghosts . 

The interaction strength between two mixed branes can be read off 
from the above formula 
\bea
T^2_{p_1,p_2} = \frac{T_{p_1}T_{p_2}}{24}\sqrt{det(1-{\cal{F}}_1)
det(1-{\cal{F}}_2)}\;\;[Tr(S_1S^{T}_2)-2]
\eea
An overall factor $\frac{1}{24}$ is used to normalize the result .

Now consider two parallel $m_p$-branes with ${\cal{F}}_1 = {\cal{F}}_2
\equiv {\cal{F}}$ in non compact spacetime, therefore equation (35) gives
\bea
{\cal{A}}_0 = 6V_{p+1}[T^2_p det(1-{\cal{F}})]G_{25-p}(Y^2)
\eea
where $G_D(Y^2)$ is the massless scalar Green's function in $D$ dimensions
and $Y^i = y^i_1-y^i_2$ is the separation of the $m_p$-branes . 

Note that result (35) can also be obtained by projecting the boundary states
onto the massless levels . Performing this projection on boundary state
$\mid B_2 \rangle$, the metric and the antisymmetric tensor components and
dilaton part of this projection for $d=26$ can be written as a state
$ \mid s_2 \rangle$ which lives in $\mid B_2 \rangle$ and has the form 
\bea
\mid s_2 \rangle &=& \frac{T_{p_2}}{2} \sqrt{det(1-{\cal{F}}_2)}\;
\delta^{(25-{p_2})}(x^{i_2} - y^{i_2}_2)(-\alpha^{\mu}_{-1}S^{(2)}_{\mu \nu} 
{\tilde \alpha}^{\nu}_{-1} 
\nonumber\\
&~& + b_{-1} {\tilde c}_{-1} +c_{-1}
{\tilde b}_{-1} ) \mid Z \rangle \mid 0 \rangle \prod_{\mu}
\mid p^{\mu}_L= p^{\mu}_R = 0 \rangle
\eea
Vanishing of all left and right components of momentum says that this state
has zero momentum and zero winding numbers . By inserting the closed
string propagator between $\mid s_1 \rangle $ (which lives in 
$\mid B_1 \rangle $ ) and $\mid s_2 \rangle $, again we obtain
the result (35) .
\section{Conclusion}
					 
We explicitly obtained boundary state for mixed boundary conditions on
 closed string and developed  the corresponding formalism to extract
 the amplitude and the contribution of the massless states on it, in presence
 of non zero back-ground $B_{\mu \nu}$ and internal gauge fields . 
 
 The formalism was applied
 to both compact and non compact spacetime . In the space with 
 compactification strong relation holds among internal windings and
 momenta restricting the total flux passing through the branes, 
 or for a given flux restricting the windings and momenta of closed string.
Only when the ratio of different total internal fluxes are fractional 
the winding modes of boundary states can have contribution to the interaction
, since they can only emitted and absorbed with certain angles restricted
by those fluxes .
 
 The formalism can be extended to include fermionic degrees of freedom 
 and hence can be applied to the superstring. This work is in progress. 

 {\bf Acknowledgement}

 The authors would like to tank M.M. Sheikh Jabbari for useful
 discussion .

\end{document}